\documentclass[a4paper,11pt]{article}
\usepackage{jcappub}
\usepackage[T1]{fontenc}
\usepackage{bm}

\title{Removing the ISW-lensing bias from the local-form primordial non-Gaussianity estimation}

\author[a,b,1]{Jaiseung Kim,\note{Corresponding author.}}
\author[c]{Aditya Rotti}
\author[a,d,e]{Eiichiro Komatsu}

\affiliation[a]{Max-Planck-Institut f\"{u}r Astrophysik, Karl-Schwarzschild Str. 1, 85741 Garching, Germany}
\affiliation[b]{Niels Bohr Institute, Blegdamsvej 17, DK-2100 Copenhagen, Denmark}
\affiliation[c]{IUCAA, Post Bag 4, Ganeshkhind, Pune-411007, India}
\affiliation[d]{Kavli Institute for the Physics and
Mathematics of the Universe, Todai Institutes for Advanced Study, the
University of Tokyo, Kashiwa, Japan 277-8583 (Kavli IPMU, WPI)}
\affiliation[e]{Texas Cosmology Center and the Department of Astronomy,
The University of Texas at Austin, 1 University Station, C1400, Austin,
TX 78712, USA}

\emailAdd{kim@mpa-garching.mpg.de}
\emailAdd{aditya@iucaa.ernet.in}
\emailAdd{komatsu@mpa-garching.mpg.de}

\abstract{The Integrated Sachs-Wolfe (ISW) effect produces a secondary
temperature anisotropy of the cosmic microwave background (CMB), as CMB
photons travel through time-varying potentials along the line-of-sight.  
The main contribution comes from redshifts $z\lesssim 2$, where dark
energy leads to a decay of potentials. As the same photons are gravitationally lensed by these decaying potentials, there exists a high degree of
correlation between the ISW effect and CMB lensing, leading to a non-zero three-point correlation (bispectrum)
of the observed temperature anisotropy. 
This ISW-lensing bispectrum, whose shape resembles that of the so-called ``local-form'' primordial bispectrum parametrized by $f_{\mathrm{NL}}$, is known to be the largest contamination of $f_{\mathrm{NL}}$. 
In order to avoid a spurious detection of primordial
non-Gaussianity, we need to remove the ISW-lensing bias. 
In this work, we investigate three debiasing methods: (I)
subtraction of an expected, ensemble average of the ISW-lensing
bispectrum; (II) subtraction of a measured ISW-lensing bispectrum; and
(III) direct subtraction of an estimated ISW signal from an observed
temperature map. One may use an estimation of the
ISW map from external non-CMB data or that from the CMB data themselves.
As the methods II and III are based on fewer
assumptions about the nature of dark energy, they are preferred over the
method I. While the methods I and II yield unbiased estimates of
$f_{\mathrm{NL}}$ with comparable error bars, the method III yields a
biased result when the underlying primordial
$f_{\mathrm{NL}}$ is non-zero and the ISW map is
estimated from a lensing potential reconstructed from the observed temperature
map. One of the sources of the bias is a lensing reconstruction noise
bias which is independent of 
$f_{\mathrm{NL}}$ and can be calculated precisely, but
other $f_{\mathrm{NL}}$-dependent terms are difficult to
compute reliably. We thus conclude that the method II is the best, model-independent way to remove the
ISW-lensing bias of $f_{\mathrm{NL}}$, enabling us to
test the physics of inflation with smaller systematic errors.
}
\begin{document}
\maketitle
\flushbottom
\section{Introduction}
Convincing detection of the so-called ``local-form'' three-point
correlation function (bispectrum) of primordial curvature perturbations
from inflation has profound implications for our understanding of the
physics of inflation, as it would rule out all single-field inflation
models \cite{singlefield_fnl_consistency,komatsu/etal:prep}, provided that an initial quantum
state of the curvature perturbation is in a preferred state called
the Bunch-Davies state \cite{agullo/parker:2011,ganc:2011} and that the
curvature perturbation does not evolve outside the horizon due to a
non-attractor solution \cite{namjoo/firouzjahi/sasaki:prep,chen/etal:prep}.\footnote{Also see workshop
summaries of ``Critical Tests of Inflation Using Non-Gaussianity'' in {\sf
http://www.mpa-garching.mpg.de/\textasciitilde{}komatsu/meetings/ng2012/}.}

The local-form bispectrum is defined as (e.g.,
\cite{komatsu/spergel:2001})
\begin{equation}
\langle\Phi_{{\mathbf k}_1}\Phi_{{\mathbf k}_2}\Phi_{{\mathbf
 k}_3}\rangle=(2\pi)^3\delta({\mathbf k}_1+{\mathbf k}_2+{\mathbf
 k}_3)(2f_{\mathrm{NL}})\left[P_\Phi(k_1)P_\Phi(k_2)+\mbox{(2
					  perm)}\right], 
\end{equation}
where $\Phi$ is Bardeen's curvature perturbation in the matter era given
by the trace of 
the space-space metric, i.e., $\sqrt{\det(g_{ij})}=a^3(t)(1+3\Phi)$, and
$a(t)$ is the Robertson-Walker scale factor. The function
$P_\Phi(k)$ is
the power spectrum of $\Phi$ defined as $\langle\Phi_{{\mathbf
k}_1}\Phi_{{\mathbf k}_2}\rangle=(2\pi)^3\delta({\mathbf k}_1+{\mathbf
k}_2)P_\Phi(k)$. The latest measurements suggest $P_\Phi(k)\propto
k^{n_s-4}$ with $n_s=0.96\pm 0.01$ (68\%~CL)
\cite{hinshaw/etal:prep,hou/etal:prep,sievers/etal:prep}. It follows
from this wavenumber-dependence of 
$P_\Phi(k)$ that the local-form bispectrum is largest in the so-called
squeezed configurations, where one of the wavenumbers is much smaller
than the other two, e.g., $k_3\ll k_1\approx k_2$
\cite{babich/creminelli/zaldarriaga:2004}. In the squeezed limit,
$k_3\to 0$, all
single-field models give $f_{\rm NL}=\frac5{12}(1-n_s)={\cal
O}(10^{-2})$ \cite{maldacena:2003,singlefield_fnl_consistency}.

The latest {\sl WMAP} 9-year limit is
$f_{\mathrm{NL}}=39.8\pm 
19.9$ (68\%~CL). The {\sl WMAP} team then subtracts $\delta f_{\rm
NL}=2.6$ from this measurement in order to correct for the
bias due to the ``ISW-lensing bispectrum,'' reporting the final limit of
$f_{\mathrm{NL}}=37.2\pm 19.9$ (68\%~CL) \cite{bennett/etal:prep}.

What is the ISW-lensing bispectrum?
The Integrated Sachs-Wolfe (ISW) effect is a secondary
temperature anisotropy caused by time-varying gravitational potential
wells between the last-scattering surface and us
\cite{ISW}. The (linear) ISW effect
vanishes during the matter era, while it becomes important
at low redshifts, $z\lesssim 2$, where dark energy leads to a
decay of potential wells. The same potential wells gravitationally
deflect the paths 
of CMB photons (see Ref.~\cite{lensing_review} for a
review). Therefore, there is a correlation between the ISW
effect, which is important only at low multipoles, $l\lesssim 10$, and
a change in CMB anisotropy due to lensing, which is
important at high multipoles, $l\gtrsim 1000$. This leads to a non-zero
bispectrum of the observed temperature anisotropy
\cite{goldberg/spergel:1999}, which is 
largest in the squeezed configuration, e.g., $l_3\ll l_1\approx l_2$
\cite{smith/zaldarriaga:2006}. Therefore, the ISW-lensing bispectrum
yields a contamination of the primordial local-form bispectrum
\cite{smith/zaldarriaga:2006,serra/cooray:2008,hanson/etal:2009,Mangilli:fnl_lensing}.
We need to properly remove the ISW-lensing bias in order to avoid a spurious detection of primordial non-Gaussianity.

What is the best way to remove the ISW-lensing bias? The
most straightforward way is to calculate the expected ISW-lensing
bispectrum given a cosmological model, and subtract it from the measured
bispectrum (``Method I''). This is what was done by the {\sl WMAP} team for the
nine-year analysis \cite{bennett/etal:prep}. However, as one can only
predict the {\it ensemble average} of the ISW-lensing bispectrum,
this method ignores a realization-dependent term. One may then
assume that the shape of the ISW-lensing bispectrum is known but the
amplitude is not, and include the amplitude of the ISW-lensing
bispectrum as a free parameter \cite{Komatsu_fnl_review}. (i.e.,
one marginalizes over the amplitude of the ISW-lensing bispectrum.)

The methods we explore in this paper go beyond these simple
treatments in two ways. For one, we first {\it measure} the ISW-lensing
cross-power spectrum from data directly, and use this measured
cross-power spectrum to compute the ISW-lensing bispectrum (``Method II''). In
this way we can fully capture the ISW-lensing bias that is actually
there in the sky. We show that, not only does this method
yield an unbiased estimate of $f_{\rm NL}$, but also yields a statistical
uncertainty in $f_{\rm NL}$ which is as small as Method I, and thus it
is optimal.

For another, we first {\it clean} the ISW effect by
removing an estimate of the ISW effect from an observed temperature map,
and then measure the bispectrum (``Method III''). To the extent that the
estimator of the 
ISW effect is accurate, this method allows us to remove the ISW-lensing
coupling before measuring the bispectrum from data. However, we find
that this method, as currently
implemented, yields a biased result, if the underlying,
primordial $f_{\rm NL}$ is non-zero and the ISW estimation
comes from a lensing potential reconstructed from the CMB
data themselves, rather than from external non-CMB data.

The outline of this paper is as follows.
In Section \ref{basics}, we briefly describe the ISW-lensing
bispectrum.  
In Section \ref{simulation}, we describe our simulations of lensed
non-Gaussian CMB temperature maps with noise.
In Section \ref{debias}, we describe three methods for removing the
ISW-lensing bias of $f_{\rm NL}$.
In Section \ref{sec:results}, we apply these methods to simulated data and
present the results. We conclude in Section \ref{conclusion}.
In Appendix \ref{quad_estimator}, we describe our estimator of the
lensing potential. In Appendix \ref{noise_bias_method2}, we derive the
estimator of $f_{\rm NL}$ for Method II.
In Appendix \ref{noise_bias}, we derive the noise bias in the reduced
bispectrum of an ISW-subtracted map of Method III, which arises
from the reconstruction noise of lensing potential.

\section{The ISW-lensing bispectrum}
\label{basics}
\subsection{Bispectrum estimator and Fisher matrix}
The CMB anisotropy measured over the whole-sky is conveniently
decomposed in terms of spherical harmonics as $T(\hat{\mathbf
n})=\sum_{lm} a_{lm}\,Y_{lm}(\hat{\mathbf n})$, 
where $\hat{\mathbf n}$ is a unit vector pointing toward a given
direction in the sky, $a_{lm}$ a decomposition coefficient, and $Y_{lm}(\hat {\mathbf n})$ a spherical harmonic function. 
The expectation value of a 3-point correlation is given by
$\langle a_{l_1m_1} a_{l_2 m_2} a_{l_3 m_3} \rangle= \mathcal G^{m_1 m_2 m_3}_{l_1 l_2 l_3}  \sum_i f^{(i)}_{\mathrm{NL}}\,b^{(i)}_{l_1 l_2 l_3}$,
where
$\langle \ldots \rangle$ denotes the ensemble average over many
realizations of universes, and $\mathcal G^{m_1 m_2 m_3}_{l_1 l_2 l_3}$
is  defined by $\mathcal G^{m_1 m_2 m_3}_{l_1 l_2
l_3} \equiv\int d^2\hat{\bf n}\:Y_{l_1 m_1}(\hat{\bf n})\:Y_{l_2
m_2}(\hat{\bf n})\:Y_{l_3 m_3}(\hat{\bf n})$. 
Here, $(i)$ denotes various sources of non-Gaussianity such
 as ``local'' and ``ISW-lensing,'' etc., 
 $b^{(i)}_{l_1 l_2 l_3}$ is the reduced bispectrum of a particular
 shape, and $f_{\rm NL}^{(i)}$ is the corresponding amplitude. See
 Ref.~\cite{komatsu/spergel:2001} for the expression of $b^{\rm 
 local}_{l_1 l_2 l_3}$ and Eq.~\eqref{isw_bispectrum} for $b^{\rm
 ISW-lensing}_{l_1 l_2 l_3}$.

Given the CMB data, we may estimate $f^{(i)}_{\mathrm{NL}}$ from
$f^{(i)}_{\mathrm{NL}}=\sum_j (\bm F^{-1})_{ij}\,S_j$,
where (see Ref.~\cite{Komatsu_fnl_review} for a review)
\begin{eqnarray}
S_i=\frac{1}{6} \sum_{lm} \mathcal G^{m_1 m_2 m_3}_{l_1 l_2 l_3} b^{(i)}_{l_1 l_2 l_3}
\left[(C^{-1} a)_{l_1 m_1}(C^{-1} a)_{l_2 m_2} (C^{-1} a)_{l_3 m_3} -3(C^{-1})_{l_1 m_1,l_2 m_2} (C^{-1}a)_{l_3 m_3}\right],\nonumber\\\label{S_i}
\end{eqnarray}
with $\mathbf C$ being the covariance matrix of data including CMB and noise, and $\bm F$ is the Fisher matrix given by 
\begin{eqnarray}
{\bm F}_{ij}&=&\frac16\sum_{lm}\sum_{l'm'}\mathcal G^{m_1 m_2 m_3}_{l_1 l_2 l_3} b^{(i)}_{l_1 l_2 l_3}
(C^{-1})_{l_1 m_1,l'_1 m'_2}\,(C^{-1})_{l_2 m_2,l'_2 m'_2}\,(C^{-1})_{l_3 m_3,l'_3 m'_3}\,b^{(j)}_{l'_1 l'_2 l'_3}\,\mathcal G^{m'_1 m'_2 m'_3}_{l'_1 l'_2 l'_3}.\nonumber\\
\end{eqnarray}
These expressions simplify greatly to those in Ref.~\cite{fnl_cubic} when the
covariance matrix is diagonal and isotropic, i.e.,
$C_{lm,l'm'}=C_l\delta_{ll'}\delta_{mm'}$. The 1$\sigma$ uncertainty in
$f_{\rm NL}^{(i)}$ is given by $\sqrt{({\bm F}^{-1})_{ii}}$.

\subsection{The bias due to the ISW-lensing bispectrum}
\label{bias_estimation}
The ISW effect is produced by
the blue-shifting and red-shifting of photons as photons fall in and
climb out of potential wells in their pathway, and is given in terms of
a time derivative of $\Psi-\Phi$ along the
line-of-sight \cite{ISW}:
\begin{eqnarray}
T_{\mathrm{ISW}}(\hat{\bf{n}})=\int^{\chi_*}_{0} d\chi[\dot\Psi-\dot\Phi](\chi \hat{\bf{n}},\eta_0-\chi),
\end{eqnarray}
where $\Psi({\bf x},\eta)$ is the Newtonian potential given by the
time-time metric, 
$g_{00}=-(1+2\Psi)$, the dot denotes a derivative with respect to
the conformal time, $\partial/\partial\eta$, $\eta_0$ is the present-day
conformal time, and $\chi^*$ is the comoving distance to the last
scattering surface.  
The ISW effect is not present during matter domination (in which
$\dot\Psi=0=\dot\Phi$), but becomes important at low redshifts,
$z\lesssim 2$, where dark energy starts to affect the evolution of
$\Psi$ and $\Phi$.  

Traveling along the line-of-sight, CMB photons are also
gravitationally lensed by the same potential as
$T(\hat{\bf{n}})\to 
T(\hat{\bf{n}}+\nabla\psi)$ (see \cite{lensing_review} for a
review). Here, $\psi$ is a ``lensing potential,'' given by
\begin{eqnarray}
\psi(\hat{\bf{n}})=-\int^{\chi_*}_{0} d\chi \frac{f_K(\chi_*-\chi)}{f_K(\chi_*)\,f_K(\chi)} [\Psi-\Phi](\chi \hat{\bf{n}},\eta_0-\chi),
\end{eqnarray}
where $f_K(\chi)$ is the comoving angular-diameter distance, which is
equal to $\chi$ in a flat universe ($K=0$).
As most of the lensing effect also comes from $z\lesssim 2$, there is a correlation between CMB anisotropy and the lensing potential.

\begin{figure}[t]
\centering
\includegraphics[width=0.5\textheight]{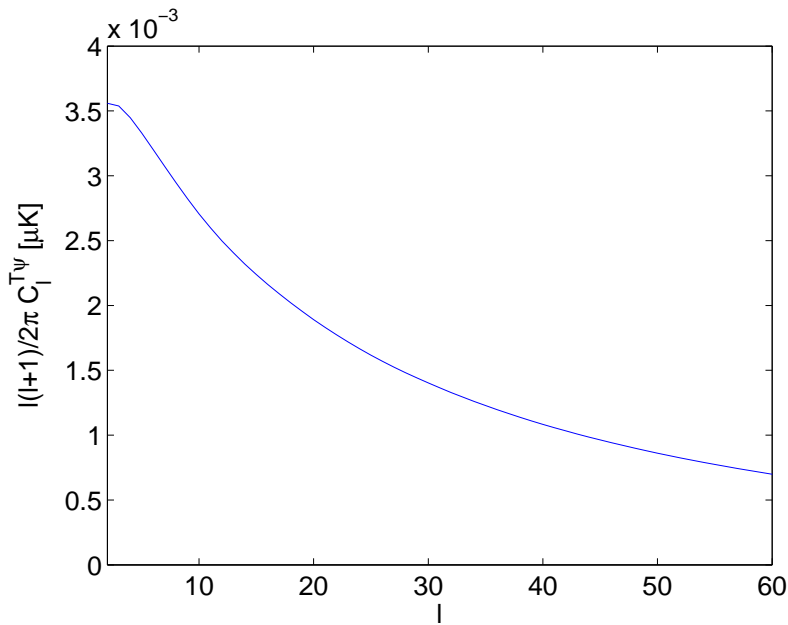}
\caption{Cross-power spectrum of CMB temperature anisotropy and the
 lensing potential, $C^{T\psi}_l$.} 
\label{Cl_Tpsi}
\end{figure}

Let us decompose the lensing potential into spherical
harmonics as $\psi_{lm}=\int d^2\mathbf{n}\;\psi(\hat{\bm
n})\,Y^*_{lm}(\hat{\bm n})$, and define the temperature-lensing
cross-power spectrum as
$C_l^{T\psi}\equiv \langle a_{lm}\psi_{lm}^*\rangle$. While this
cross-power spectrum at low multipoles, $l\lesssim 20$, is almost
entirely dominated by the above late ISW-lensing correlation, 
there are also other small contributions (see Fig.~4 of
Ref.~\cite{lewis:2012}). We use the CAMB code \cite{CAMB} to include
these smaller effects as well. 
The linear theory calculation is an excellent approximation, as non-linear effects have no impact on the contamination of $f_{\rm NL}$ \cite{junk/komatsu:2012}.
In Fig. \ref{Cl_Tpsi}, we show the cross-correlation power spectrum,
$C^{T\psi}_l$, for the best-fit $\Lambda$CDM parameters given by the ``{\sl
WMAP}7+BAO+$H_0$'' combination in Ref.~\cite{WMAP7:Cosmology}. 

The temperature-lensing correlation generates the following
bispectrum even in the absence of primordial non-Gaussianity
\cite{goldberg/spergel:1999}: 
\begin{eqnarray}
\left.\langle a_{l_1 m_1}\,a_{l_2 m_2}\,a_{l_3 m_3}\rangle\right|_{f_{\mathrm{NL}}=0}= \mathcal G^{m_1 m_2 m_3}_{l_1 l_2 l_3} b^{\mathrm{ISW-lensing}}_{l_1 l_2 l_3},
\end{eqnarray}
where
\begin{eqnarray}
b^{\mathrm{ISW-lensing}}_{l_1 l_2 l_3}= \frac{l_2(l_2+1)+l_3(l_3+1)-l_1(l_1+1)}{2}C^{TT}_{l_2}C^{T\Psi}_{l_3} + (5\;\mathrm{perm}),\label{isw_bispectrum}
 \end{eqnarray}
where $C^{TT}_{l}$ is the {\it lensed} (rather than unlensed, as
pointed out by \cite{Shape_lensing_bispectrum}) power spectrum of CMB
temperature anisotropy.
 As $C_l^{T\psi}$ falls off rapidly with multipoles (see
Fig.~\ref{Cl_Tpsi}), this bispectrum is largest in the squeezed
configurations, e.g., $l_3\ll l_1\approx l_2$, just like the local-form
bispectrum. This is the reason why the ISW-lensing bispectrum results in a
contamination of $f_{\rm NL}$. The expected bias in $f_{\rm NL}^{\rm
local}$ due to the ISW-lensing coupling, $\delta f_{\rm NL}$, can be computed from
\begin{eqnarray}
\nonumber
& &\delta f_{\rm NL}=\\
& &\frac{\sum_{lm}\sum_{l'm'}\mathcal G^{m_1 m_2 m_3}_{l_1 l_2 l_3}
 b^{\rm local}_{l_1 l_2 l_3}
(C^{-1})_{l_1 m_1,l'_1 m'_2}\,(C^{-1})_{l_2 m_2,l'_2 m'_2}\,(C^{-1})_{l_3 m_3,l'_3 m'_3}\,b^{\mathrm{ISW-lensing}}_{l'_1 l'_2 l'_3}\,\mathcal G^{m'_1 m'_2 m'_3}_{l'_1 l'_2 l'_3}}{\sum_{lm}\sum_{l'm'}\mathcal G^{m_1 m_2 m_3}_{l_1 l_2 l_3} b^{\rm local}_{l_1 l_2 l_3}
(C^{-1})_{l_1 m_1,l'_1 m'_2}\,(C^{-1})_{l_2 m_2,l'_2 m'_2}\,(C^{-1})_{l_3 m_3,l'_3 m'_3}\,b^{\rm local}_{l'_1 l'_2 l'_3}\,\mathcal G^{m'_1 m'_2 m'_3}_{l'_1 l'_2 l'_3}}.\nonumber\\\label{bias}
\end{eqnarray}
In Table \ref{observation_bias}, we show the expected bias and the
corresponding 1$\sigma$ uncertainty in $f_{\rm NL}$ for {\sl
Planck} as well as for a 
cosmic-variance-limited experiment measuring temperature anisotropy up
to $l=2500$ (no polarization information is used).
The ISW-lensing bias exceeds the expected $1\sigma$ uncertainty of the
upcoming {\sl Planck data}, and thus it must be removed. 
For a cosmic-variance-limited experiment, the expected bias is four
times the $1\sigma$ uncertainty. 

\begin{table}[t]
\centering
\begin{tabular}{|c|cc|}
\hline
  & 1$\sigma$ error & bias \\
\hline
Planck sensitivity &  5.1 &  7.8\\
Cosmic-variance-limited ($l\le 2500$) & 3.3 & 13.0\\
\hline
\end{tabular}
\caption{\label{observation_bias} The expected bias and the corresponding 1$\sigma$ uncertainty of
 $f_{\mathrm{NL}}$ for a {\sl Planck}-like experiment and a
 cosmic-variance-limited 
 experiment measuring temperature 
 anisotropy up to $l=2500$. The uncertainty presented here does not include the statistical fluctuation from the ISW-lensing bias.}
\end{table}

\section{Simulation}
\label{simulation}
To test validity of our methods for removing the ISW-lensing
bias of $f_{\rm NL}$ described in the next section, we apply our methods
to simulated lensed non-Gaussian temperature maps with noise.
We use 1000 simulated {\it unlensed} non-Gaussian temperature maps
produced by Elsner 
and Wandelt \cite{fnl_sim}. The cosmological parameters of the simulations are: $\Omega_\Lambda=0.728$, $\Omega_ch^2=0.1123$,
$\Omega_bh^2=0.0226$, $h=0.704$, $n_s=0.963$, $\tau=0.087$, and
$\Delta_{\cal R}^2(k_0)=2.441\times 10^{-9}$ with $k_0=0.002~{\rm
Mpc}^{-1}$. These simulations provide a Gaussian piece, $a_{lm}^{\rm
L}$, and a non-Gaussian piece, $a^{\mathrm{NL}}_{lm}$, for $l\le
3500$. The total anisotropy is then given by
$a_{lm}=a^{\mathrm{L}}_{lm} 
+f_{\mathrm{NL}}\,a^{\mathrm{NL}}_{lm}$. 

\begin{figure}[t]
\centering
\includegraphics[width=0.33\textheight]{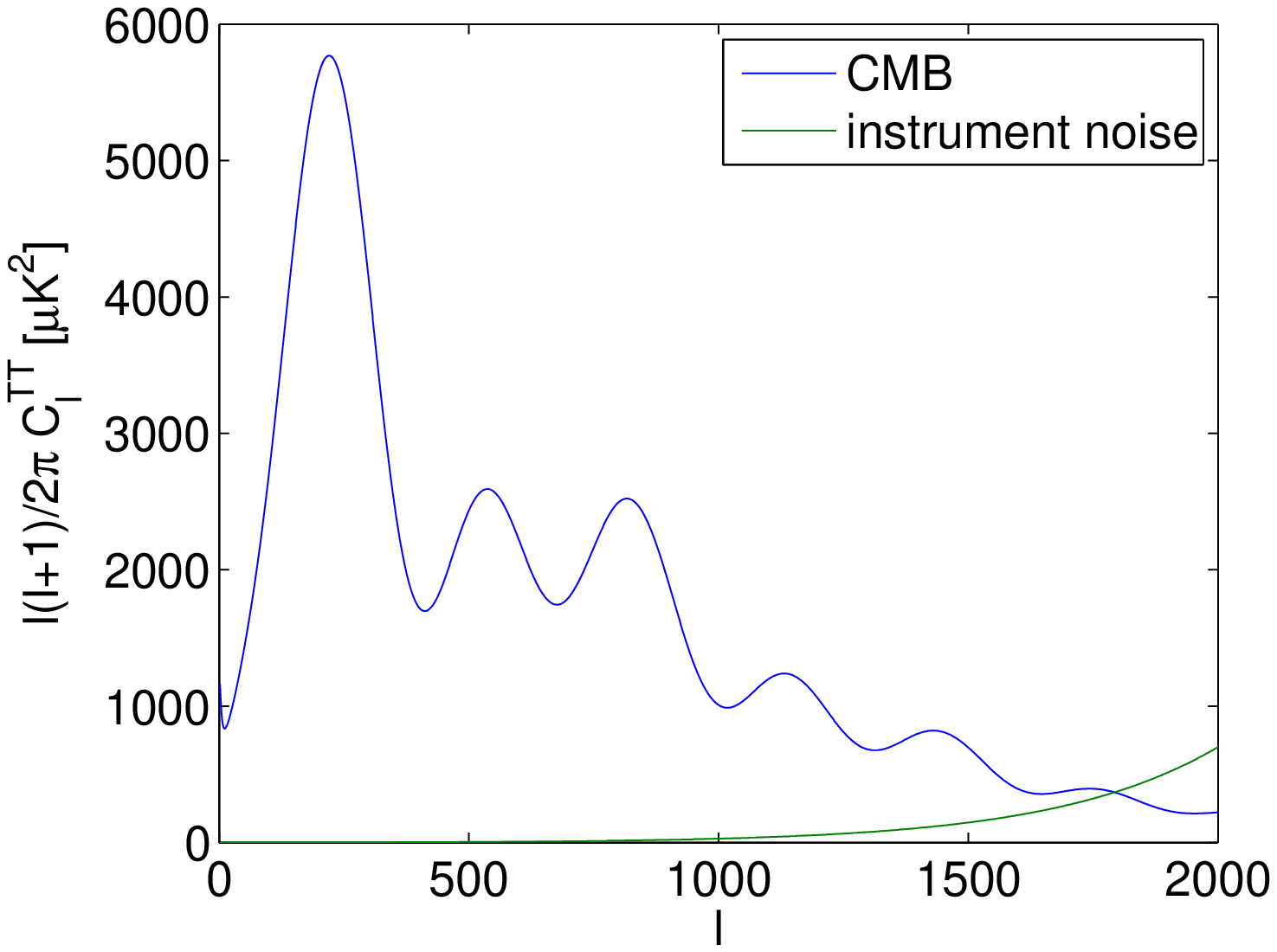}
\includegraphics[width=0.33\textheight]{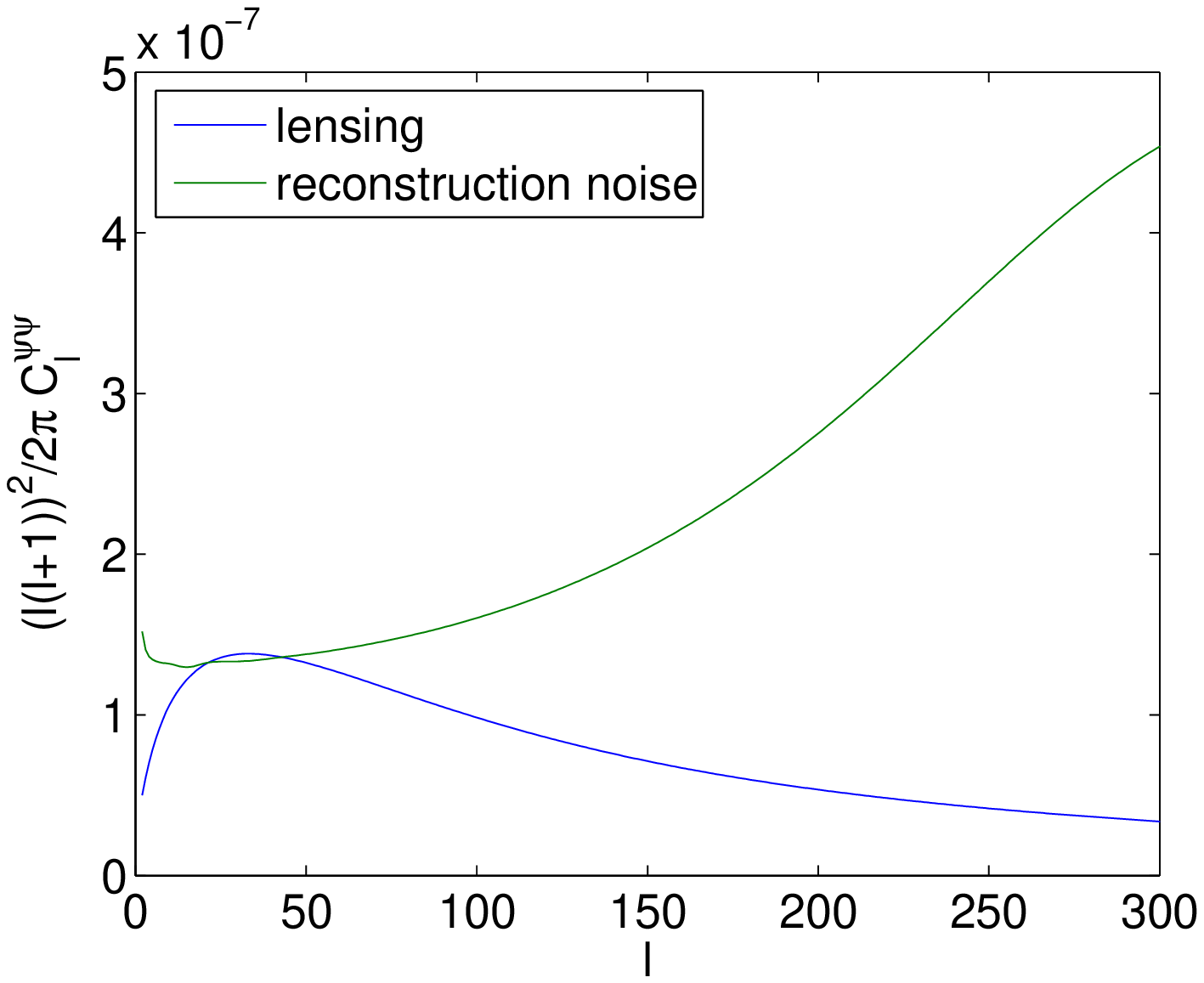}
\caption{(Left) Power spectra of CMB and the expected {\sl Planck}
 noise bias. (Right) Power spectra of the lensing potential and the expected
 {\sl Planck} reconstruction noise bias (computed from Eq.~\eqref{Nl_recon}).}
\label{noise}
\end{figure}

As the lensing potential is not available from these
simulations, we need to generate the lensing potential such that it has
a proper correlation with the pre-generated total
CMB anisotropy, $a_{lm}$.
In order to do this, we use a ``constrained Gaussian realization''
method \cite{Constrained_Gaussian_Simple,Constrained_Gaussian,harmonic_inpainting}.
We have checked that the correlation between the simulated lensing
potential and $a_{lm}$ agrees with the theoretical expectation.
We then use the LensPix code \cite{LensPix} to 
lens the simulated CMB maps with the lensing
potential we have generated. Finally, to these lensed CMB maps we add Gaussian,
white, and homogeneous noise with a given noise power
spectrum. (For simplicity we do not include inhomogeneity of {\sl
Planck} noise.)
We use the FUTURCMB code \cite{FUTURCMB} to calculate the noise power spectrum corresponding to the
expected sensitivity of {\sl Planck} \cite{Planck_bluebook}.
In the left panel of Fig. \ref{noise}, we show the noise power spectrum
together with the CMB temperature power spectrum.

Once maps are generated, we use the method of Ref.~\cite{fnl_cubic} to estimate
$f_{\mathrm{NL}}$ from full-sky maps
(i.e., no mask is applied). The functions
required for computing the local-form CMB bispectrum, $\alpha(r)$ and
$\beta(r)$ \cite{komatsu/spergel:2001,fnl_cubic}, are computed by the CAMB 
code \cite{CAMB} with 1954 points in the radial coordinates, $r$.
These are chosen in accordance with CAMB's internal $k$
sampling, which are due to the $j_l(kr)$ terms in the integrand for the $\alpha(r)$ and
$\beta(r)$.

\begin{table}[t]
\centering
\begin{tabular}{|c|c|c|}
\hline
$f_{\mathrm{NL}}$ & Planck Sensitivity & Cosmic Variance Limited ($l\le 2500$) \\
\hline
0  & $7.6 \pm 5.1$ &$12.8 \pm 3.9$\\
20  & $27.7 \pm 5.7$ & $ 33.1 \pm 4.9$\\
40  & $47.7 \pm  7.1$ & $53.3 \pm  6.7$\\
\hline
\end{tabular}
\caption{\label{sim0} Mean and standard deviation of
 $f_{\mathrm{NL}}$ estimated from 1000 simulated lensed
 CMB maps with {\sl Planck}-like noise (the second column) and with no
 noise (the third column) for the input values of
 $f_{\mathrm{NL}}=0$, 20, and 40.}
\end{table}

In Table \ref{sim0}, we show the average and standard
deviation of $f_{\mathrm{NL}}$ estimated from 1000
lensed CMB plus {\sl Planck}-like noise simulations (the second column)
as well as from cosmic-variance-limited simulations (the third column)
before removing the ISW-lensing bias, for the input values of
$f_{\mathrm{NL}}=0$, 20, and 40.  
For all cases, the biases we find agree with the expectations given in
Table~\ref{observation_bias}.  
The standard deviation increases for larger values of
$f_{\mathrm{NL}}$ due to the contribution of
non-Gaussian terms to the covariance matrix of the bispectrum.
As $f_{\rm NL}$ increases, the error bars become more dominated 
by non-Gaussian contributions to the covariance matrix. Therefore a
Gaussian piece, which includes instrumental noise, becomes less
important.

\section{Removing the ISW-lensing bias}
\label{debias}
In this section, we describe three methods for removing the ISW-lensing bias. 
\subsection{Fitting out the ISW-lensing bispectrum (Method Ia and Ib)}
\label{debias1}
The simplest possible method to remove the bias is to
subtract the ensemble average of the ISW-lensing
bias given by Eq.~\eqref{bias} from the measured
$f_{\mathrm{NL}}$ (Method Ia).
While this is the simplest method, it comes with a couple of
caveats. First, it assumes a perfect knowledge of the ISW-lensing
bispectrum. While this assumption is not too unreasonable given the
success of the minimal $\Lambda$CDM model, it might still be too restrictive
given the fact that we do not know the precise nature of dark energy. Second,
even if dark energy is a cosmological constant, one can only predict the
ensemble average of the ISW-lensing bispectrum, whereas the bias in
$f_{\rm NL}$ is caused by a particular realization of potentials in our
past light cone.  

To partially mitigate both issues, one may simultaneously
fit $f_{\mathrm{NL}}$ and the overall amplitude of the ISW-lensing
bispectrum \cite{Komatsu_fnl_review} (Method Ib). This is
equivalent to marginalizing over the amplitude of the ISW-lensing bispectrum, i.e., we
assume that the shape of the ISW-lensing bispectrum is known precisely,
but the amplitude is not.

\subsection{Realization-dependent debiasing (Method II)}
\label{debias2}
Can we do better? The answer is yes, if we have information
on the lensing potential. How do we obtain information on the lensing
potential? One way is to use deep galaxy survey data (e.g., Euclid
\cite{Euclid_study,Euclid_science}) to measure the lensing potential
totally independent of the CMB data, and another way is to use non-Gaussian
signatures of CMB temperature anisotropy caused by gravitational lensing
to reconstruct the lensing potential. (For discussion on practical
applications of the lensing reconstruction technique to the {\sl Planck}
data, see 
\cite{lensing_recon_Planck,lensing_recon_inhomogeneous}).
Throughout this paper, we shall use the lensing potential reconstructed
from the temperature data.

Once we obtain a map of the lensing potential, we can
remove the ISW-lensing bias in two ways. The first method, which we
shall call ``Method II,'' is to estimate the ISW-lensing bispectrum:
\begin{eqnarray}
\hat b^{\mathrm{ISW-lensing}}_{l_1 l_2
 l_3}=\frac{l_2(l_2+1)+l_3(l_3+1)-l_1(l_1+1)}2 C^{TT}_{l_2}\hat
 C^{T\Psi}_{l_3}+ (5\;\mathrm{perm}),\label{lensing_bl} 
\end{eqnarray}
where
$C^{TT}_l$ is the theoretical temperature power spectrum and $\hat
C^{T\Psi}_l=(2l+1)^{-1}\sum^{l}\limits_{m=-l} a_{lm}\,\hat \psi^*_{lm}$ 
is the temperature-lensing
cross-power spectrum estimated from the data. Here,
$\hat \psi_{lm}$ is a lensing potential reconstructed from the CMB data.
Using Eq. \eqref{quad_estimator_eq} and \eqref{noise_recon}, one can
show that the expectation value of $C^{TT}_{l_2}\hat
C^{T\Psi}_{l_3}$  is given by (see Sec.~\ref{noise_bias_method2} for details):
\begin{eqnarray}
\langle C^{TT}_{l_2}\hat
 C^{T\Psi}_{l_3}  \rangle =  C^{TT}_{l_2} C^{T\Psi}_{l_3} + f_{\rm NL}\,C^{TT}_{l_2} \xi_{l_3},
\end{eqnarray}
where
\begin{eqnarray}
\xi_L\equiv \frac{N_L}{2L+1}\sum_{l_1'l_2'}\sqrt{\frac{(2l'_1+1)(2l'_2+1)(2L+1)}{4\pi}} \left(\begin{array}{ccc}l_1'&l_2'&L\\0&0&0\end{array}\right)\left(C^{TT}_{l_1'}F_{l_2'Ll_2'}+C^{TT}_{l_2'}F_{l_1'Ll_2'}\right)\frac{b^{\rm local}_{l'_1l'_2L}}{2C_{l_1'}^{tot}C_{l_2'}^{tot}},\nonumber\\
\end{eqnarray}
and $C_{l}^{tot}$ is the sum of $C^{TT}_l$ and the noise power
spectrum. Noting this result, we obtain the unbiased estimator of $f_{\rm NL}$ as
\begin{eqnarray}
\lefteqn{\hat{f}_{\rm NL}}\label{bias_obs}\\
&=&\frac{\tilde S_{\rm local}-\frac{1}{6}\sum\limits_{lm}\sum\limits_{l'm'}\mathcal G^{m_1 m_2 m_3}_{l_1 l_2 l_3} \tilde b^{\rm local}_{l_1 l_2 l_3}
(C^{-1})_{l_1 m_1,l'_1 m'_2}\,(C^{-1})_{l_2 m_2,l'_2 m'_2}\,(C^{-1})_{l_3 m_3,l'_3 m'_3}\,\hat b^{\mathrm{ISW-lensing}}_{l'_1 l'_2 l'_3}\,\mathcal G^{m'_1 m'_2 m'_3}_{l'_1 l'_2 l'_3}}{\frac{1}{6}\sum\limits_{lm}\sum\limits_{l'm'}\mathcal G^{m_1 m_2 m_3}_{l_1 l_2 l_3} \tilde b^{\rm local}_{l_1 l_2 l_3}
(C^{-1})_{l_1 m_1,l'_1 m'_2}\,(C^{-1})_{l_2 m_2,l'_2 m'_2}\,(C^{-1})_{l_3 m_3,l'_3 m'_3}\,\tilde b^{\rm local}_{l'_1 l'_2 l'_3}\,\,\mathcal G^{m'_1 m'_2 m'_3}_{l'_1 l'_2 l'_3}},\nonumber
\end{eqnarray}
where 
\begin{eqnarray}
\tilde S_{\rm local}\equiv\frac{1}{6} \sum_{lm} \mathcal G^{m_1 m_2 m_3}_{l_1 l_2 l_3} \tilde b^{\rm local}_{l_1 l_2 l_3}
\left[(C^{-1} a)_{l_1 m_1}(C^{-1} a)_{l_2 m_2} (C^{-1} a)_{l_3 m_3} -3(C^{-1})_{l_1 m_1,l_2 m_2} (C^{-1}a)_{l_3 m_3}\right],\nonumber\\
\end{eqnarray}
and 
\begin{eqnarray}
\tilde b^{\rm local}_{l_1 l_2 l_3}\equiv b^{\rm local}_{l_1 l_2 l_3}-\left(
\frac{l_2(l_2+1)+l_3(l_3+1)-l_1(l_1+1)}2 C^{TT}_{l_2} \xi_{l_3}+ (5\;\mathrm{perm})\right).
\end{eqnarray}

\subsection{Subtracting ISW from an observed temperature map
  (Method III)} 
\label{debias3}
The second method, which we shall call ``Method III,'' is to
estimate the ISW effect in our sky from a map of the lensing potential,
and subtract it from an observed temperature map. To the extent that
the estimated ISW effect is accurate, this ISW-subtracted map should
yield a vanishing ISW-lensing bispectrum. This is indeed possible, as
the lensing potential and ISW effect are highly correlated. The harmonic
coefficients of an ISW-subtracted map, $\check a_{lm}$, are given by
\cite{fnl_lensing_tracers}
\begin{eqnarray}
\check a_{lm}=a_{lm} - \frac{C^{T\psi}_l}{C^{\psi\psi}_l}\, \hat \psi_{lm}, \label{check_alm}
\end{eqnarray}
where $\hat \psi_{lm}$ is the lensing potential estimated from data, and 
$C^{T\psi}_l$ and $C^{\psi\psi}_l$ are calculated from a given
cosmological model. The power spectrum of $\check a_{lm}$ is given by
$C_l^{TT}-(C_l^{T\psi})^2/C_l^{\psi\psi}$, and the cross-correlation between
$\check a_{lm}$ and the lensing potential vanishes: $\langle\,\check
a_{lm}\:\psi^*_{lm}\rangle=0$.

\section{Results}
\label{sec:results}
The results of Method Ia and Ib are shown in Table~\ref{sim1a} and
\ref{sim1b}, respectively. Both methods remove the bias successfully,
while Method Ib (which marginalizes over the amplitude of the
ISW-lensing bispectrum) yields slightly larger uncertainties in
$f_{\mathrm{NL}}$. 

In order to apply Method II and III, we need a map of the
lensing potential. In this paper we estimate a map of the lensing
potential from simulated CMB temperature maps using the lensing
reconstruction technique of 
Ref.~\cite{CMB_Lensing:full-sky} with the unlensed CMB power spectrum in the
filter replaced by
the lensed CMB power spectrum to account for higher-order terms in
the lensing potential 
\cite{Shape_lensing_bispectrum}. See Appendix \ref{quad_estimator} for
details of our estimator. 

The results of Method II are shown in Table \ref{sim2}. 
We find that the uncertainties in
$f_{\mathrm{NL}}$ from Method II are quite comparable to those of Method
I. Method II is superior to Method I, as it is based on fewer assumptions
about the nature of dark energy than Method I while keeping optimality of
the estimator. We thus recommend Method II
as the best method to remove the ISW-lensing bias.

Finally, the results of Method III are shown in Table
\ref{sim3}: $f_{\mathrm{NL}}$ is estimated from
ISW-subtracted maps given by Eq.~\eqref{check_alm}.
There is one subtlety in this method. As a map of the lensing
potential is reconstructed from CMB data themselves and the
estimated $\psi_{lm}$ is given by a product of two $a_{lm}$'s, there is a
non-zero three-point correlation between $a_{lm}$ and 
the reconstruction error (i.e., the
difference between the true $\psi_{lm}$ and the reconstructed one): $\langle
a_{l_1m_1}a_{l_2m_2}(\hat{\psi}_{l_3m_3}-\psi_{l_3m_3})\rangle\ne
0$. This correlation produces the noise bias in the bispectrum measured
from ISW-subtracted maps. One can calculate and subtract this
noise bias. We derive the formula for the noise bias in Appendix
\ref{noise_bias}. We find that the noise bias in $f_{\rm NL}$ is $\delta
f_{\rm NL}= -25.3$ for the {\sl Planck} noise level, and $\delta f_{\rm
NL}=-8.1$ for the cosmic-variance-limited case. These biases have been
subtracted already in the values quoted in Table \ref{sim3}.

However, as the lensing reconstruction relies on
non-Gaussian signatures of temperature anisotropy induced by lensing, the
presence of primordial non-Gaussianity yields a small but non-negligible
impact on the reconstructed lensing potential, especially on large
angular scales \cite{merkel/schaefer:prep}. Specifically, the lensing
estimator uses the fact that the cross-power spectrum between different
multipoles, $\langle a_{lm}a_{l'm'}\rangle$, is correlated with a
long-wavelength lensing potential, $\psi_{LM}$.  This is similar 
to what the local-form bispectrum does: $\langle a_{lm}a_{l'm'}\rangle$
is correlated with a long-wavelength mode, $a_{LM}$. The lensing
correlation peaks at $L\approx 
50$, whereas the local form peaks at $L=2$.  As a result, the
reconstructed lensing potential map at small $L$ receives contributions from
$f_{\mathrm{NL}}$-dependent terms, giving a bias in $f_{\rm NL}$. 
Unlike Method II, for which we are able to calculate the $f_{\rm
NL}$-dependent terms accurately (see Appendix~\ref{noise_bias_method2}),   
the $f_{\mathrm{NL}}$-dependent terms in the estimator of Method III are
difficult to compute reliably. We thus conclude that Method III, as
currently implemented, yields a biased result, if the ISW is estimated
from a lensing potential 
reconstructed from the observed CMB temperature data themselves, and the
underlying primordial $f_{\rm NL}$ is non-zero.

\begin{table}[t]
\centering
\begin{tabular}{|c|c|c|c|}
\hline
$f_{\mathrm{NL}}$ & Planck Sensitivity& Cosmic Variance Limited ($l\le 2500$)\\
\hline
0  & $-0.15 \pm 5.1$  & $-0.13 \pm  3.9$\\
20  & $19.9 \pm 5.7$ & $20.1 \pm 4.9$\\
40  & $40.0 \pm 7.1$ & $40.3 \pm 6.7$\\
\hline
\end{tabular}
\caption{\label{sim1a} Method Ia: subtraction of the predicted ISW-lensing bispectrum.}
\end{table}
\begin{table}[t]
\centering
\begin{tabular}{|c|c|c|c|}
\hline
$f_{\mathrm{NL}}$ & Planck Sensitivity& Cosmic Variance Limited ($l\le 2500$)\\
\hline
0  & $-0.03\pm 5.3$  & $0.13 \pm 4.0$\\
20  & $20.0 \pm  5.9$ & $20.3 \pm 5.0$\\
40  & $ 40.1 \pm  7.4$ & $40.5 \pm 7.0$\\
\hline
\end{tabular}
\caption{\label{sim1b} Method Ib: the overall amplitude of the ISW-lensing bispectrum is fitted simultaneously with $f_{\mathrm{NL}}$.}
\end{table}
\begin{table}[t]
\centering
\begin{tabular}{|c|c|c|}
\hline
$f_{\mathrm{NL}}$ &  Planck Sensitivity& Cosmic Variance Limited ($l\le 2500$)\\
\hline
0  & $0.03 \pm  5.7$ &  $ 0.35 \pm 4.8$ \\
20  & $20.1 \pm  6.1$ & $ 20.5 \pm 5.3$\\
40  & $40.1 \pm  7.4$ & $ 40.7 \pm  6.7$\\
\hline
\end{tabular} 
\caption{\label{sim2} Method II: the ISW-lensing bispectrum is computed from the {\it measured} temperature-lensing cross-power
 spectrum.}
\end{table}

\begin{table}[t]
\centering
\begin{tabular}{|c|c|c|}
\hline
$f_{\mathrm{NL}}$ & Planck Sensitivity& Cosmic Variance Limited  ($l\le 2500$)\\
\hline
0  & $-0.44 \pm  5.8$ &  $-0.35 \pm 4.0$ \\
20  & $16.8 \pm 5.8$ & $16.0 \pm 4.1$\\
40  & $30.6 \pm 6.4$ & $ 27.3 \pm 4.8$\\
\hline
\end{tabular}
\caption{\label{sim3} Method III: $f_{\mathrm{NL}}$ is
 estimated from ISW-subtracted temperature maps given by Eq.~\eqref{check_alm}. The lensing reconstruction noise bias has been subtracted.}
\end{table}

\section{Conclusion}
\label{conclusion}
The ISW-lensing bispectrum, whose shape is similar to that of the local-form primordial bispectrum, biases the estimation of $f_{\mathrm{NL}}$.
For a {\sl Planck}-like experiment and a cosmic-variance-limited
experiment measuring the temperature anisotropy up to $l=2500$, we
expect the bias on $f_{\mathrm{NL}}$  to be  $7.8$ and $13$,
respectively. In order to avoid a spurious detection of the local-form
primordial bispectrum,  we must remove this ISW-lensing bias.

The method used by the {\sl WMAP} team \cite{bennett/etal:prep} assumes
that we have a perfect 
knowledge of the ISW-lensing bispectrum (Method Ia). One can relax this
assumption by marginalizing over the amplitude of the ISW-lensing
bispectrum (Method Ib). While these methods remove the ISW-lensing bias
in $f_{\rm NL}$ successfully, they rely on the assumption that we understand
the precise nature of dark energy.  

Moreover, what produces the bias
in $f_{\rm NL}$ is the ISW-lensing correlation in {\it our sky}, rather
than the ensemble average of the ISW-lensing correlation. Therefore, a
better method is to use the {\it measured} ISW-lensing correlation to
compute the ISW-lensing bispectrum and subtract it from the measured
bispectrum (Method II). We find that this method also successfully
eliminates the bias in $f_{\rm NL}$, and yields statistical
uncertainties which are as small as those of Method I. Therefore, not
only is Method II model-independent, but it is also optimal.

Another method, which removes an estimate of the ISW effect directly
from a map (Method III), is also promising, provided that the ISW
estimation comes 
from external, non-CMB data, such as galaxy surveys. However, if the ISW is
estimated from a lensing potential reconstructed from the CMB data
themselves, then $f_{\rm NL}$ estimated from the ISW-subtracted map is
biased in two ways: (1) the lensing reconstruction noise produces a
noise bias in the bispectrum of the ISW-subtracted map; and (2) the
presence of primordial $f_{\rm NL}$ biases the lensing potential
reconstruction \cite{merkel/schaefer:prep}. 
While the former effect is precisely calculable, the latter effect is difficult to estimate
reliably.

Nevertheless, it may be worth pursuing Method III further,
as removing the ISW from the observed temperature map has an added
benefit. While the standard estimator derived by Ref.~\cite{fnl_cubic} is
optimal when the underlying $f_{\rm NL}$ is zero, it becomes sub-optimal
when non-zero 
$f_{\rm NL}$ is detected with high statistical significance. A method to
make the estimator optimal even in the case of high signal-to-noise ratio
detection of $f_{\rm NL}$ relies on our knowledge of large-scale
temperature anisotropy {\it at the decoupling epoch}
\cite{improved_fnl_estimator_flat,pdf_fnl_sim}; however, this
information cannot be extracted precisely due to the presence of the ISW
effect in a low-redshift universe. Therefore, one can achieve a smaller
statistical uncertainty on 
$f_{\rm NL}$, if the ISW effect can be removed from the temperature
map \cite{improved_fnl_estimator}. This is precisely what we have
attempted to do in this paper, but the ISW signal estimated from the
lensing potential reconstructed from the temperature data was biased due
to the presence of $f_{\rm NL}$ affecting the lensing reconstruction.
Whether one can mitigate this issue by using, e.g.,
lensing potential reconstructed only from polarization data; an improved
(perhaps iterative) estimator of the lensing potential in the presence
of $f_{\rm NL}$; etc, remains to be seen.

In summary, we regard Method II as the best,
model-independent way to remove the 
ISW-lensing bias in $f_{\rm NL}$ from the forthcoming {\sl Planck} data
as well as from cosmic-variance-limited data. \footnote{After this paper was submitted, the Planck collaboration reported constraints on $f_{\rm NL}$ \cite{planck2013fnl}. They use Method Ib to remove the ISW-lensing bias.}

\acknowledgments
We thank F. Elsner and B. D. Wandelt for making their simulated
temperature maps with the local-form non-Gaussianity publicly available
\cite{fnl_sim}. We acknowledge use of CAMB \cite{CAMB}, FUTURCMB
\cite{FUTURCMB}, HEALPix \cite{HEALPix:Primer,HEALPix:framework}, and
LensPix \cite{LensPix}. 

\appendix
\section{Quadratic estimator of lensing potential}
\label{quad_estimator}

An estimate of the lensing potential in harmonic space,
$\hat{\psi}_{LM}$, may be reconstructed by the quadratic estimator as
follows \cite{CMB_Lensing:full-sky,Shape_lensing_bispectrum}: 
\begin{equation}
 \hat{\psi}_{LM}=N_L\sum_{l_1'm_1'}\sum_{l_2'm_2'}(-1)^M\left(\begin{array}{ccc}l_1'&l_2'&L\\m_1'&m_2'&-M\end{array}\right)\left(C_{l_1'}F_{l_2'Ll_1'}+C_{l_2'}F_{l_1'Ll_2'}\right)\frac{a_{l_1'm_1'}a_{l_2'm_2'}}{2C_{l_1'}^{tot}C_{l_2'}^{tot}}, \label{quad_estimator_eq} 
\end{equation}
where $C_l$ is the power spectrum of the lensed CMB (without noise), and
\begin{eqnarray}
N_L\equiv\frac{2L+1}{\sum_{l'_1l'_2}\frac{(C_{l_1'}F_{l_2'Ll_1'}+C_{l_2'}F_{l_1'Ll_2'})^2}{2C_{l_1'}^{tot}C_{l_2'}^{tot}}}, \label{Nl_recon}
\end{eqnarray}
\begin{eqnarray}
 F_{l_1Ll_2}&\equiv& \frac{L(L+1)+l_2(l_2+1)-l_1(l_1+1)}{2}I_{l_1Ll_2},\label{Flll}\\
 I_{l_1Ll_2}&\equiv& \sqrt{\frac{(2l_1+1)(2L+1)(2l_2+1)}{4\pi}}
\left(\begin{array}{ccc}l_1&L&l_2\\0&0&0\end{array}\right).\label{Illl}
\end{eqnarray}
Here, $C_{l}^{tot}$ is the sum of $C_l$ and the noise power
spectrum. One can show that the power spectrum of the reconstruction
noise bias is equal to $N_L$ \cite{CMB_Lensing:full-sky}. 
We can rewrite Eq.~\eqref{quad_estimator_eq} into the form that can be computed more efficiently:
\begin{equation}
\hat{\psi}_{LM}=\frac{1}{2} \int d^2 \hat{\bm{n}}\,\left[L(L+1)A(\hat{\bm{n}})\,B(\hat{\bm{n}})+A(\hat{\bm{n}})\,\tilde B(\hat{\bm{n}})-\tilde A(\hat{\bm{n}})\,B(\hat{\bm{n}})\right]
Y^*_{LM}(\hat{\bm{n}}),\label{quad_real_eq}
\end{equation} 
where
\begin{equation}
A(\hat{\bm{n}})\equiv\sum_{lm}
 \frac{a_{lm}}{C_{l}^{tot}}\,Y_{lm}(\hat{\bm{n}}),\qquad
B(\hat{\bm{n}})\equiv\sum_{lm} \frac{C_l\,a_{lm}}{C_{l}^{tot}}\,Y_{lm}(\hat{\bm{n}}),\label{eq:ab}
\end{equation}
\begin{equation}
\tilde A(\hat{\bm{n}})\equiv\sum_{lm}
 \frac{l(l+1)\,a_{lm}}{C_{l}^{tot}}\,Y_{lm}(\hat{\bm{n}}),\qquad
\tilde B(\hat{\bm{n}})\equiv\sum_{lm} \frac{l(l+1)C_l\,a_{lm}}{C_{l}^{tot}}\,Y_{lm}(\hat{\bm{n}}).\label{eq:abtilde}
\end{equation}
Using the HEALPix code \cite{HEALPix:Primer,HEALPix:framework}, forward
and backward spherical harmonic transformation necessary for the
equation above can be done efficiently. 
Alternatively, Eq. \eqref{quad_real_eq} can be expressed in the integral
form involving the gradient of spherical harmonics
\cite{CMB_Lensing:full-sky,lensing_recon_Planck}. 

Given the finite pixel size of the map we use for this real-space
estimator, we find that accuracy of the lensing reconstruction of the lowest
multipoles ($L\lesssim 10$), which relies on the highest multipoles
available in the temperature map,\footnote{Eq.~\eqref{quad_real_eq}
shows that the lensing potential estimator is dominated by the second
and third terms when $L$ is small. These terms contain $\tilde
A(\hat{\bm{n}})$ and $\tilde B(\hat{\bm{n}})$, which have an extra factor of
$l(l+1)$ (see Eq.~\eqref{eq:abtilde}) and thus are  more sensitive to
the finite pixel effect.}
is compromised by the finite pixel effect,
even when we use the highest resolution of HEALPix ($N_{\rm side}=8192$). 
Therefore, we use the harmonic-space estimator
(Eq.~\eqref{quad_estimator_eq}) for the reconstruction of low multipoles
($L\le 30$) and the real-space estimator (Eq.~\eqref{quad_real_eq}) for
the reconstruction of high multipoles ($30<L\le
2500$). 

Finally, difference between the true lensing potential,
$\psi_{LM}$, and an estimated one, $\hat\psi_{LM}$, is given by
\begin{eqnarray}
\nonumber
\lefteqn{n_{LM}\equiv \hat\psi_{LM}-\psi_{LM}}\label{noise_recon}\\ &=&N_L\sum_{l_1'm_1'}\sum_{l_2'm_2'}(-1)^M\left(\begin{array}{ccc}l_1'&l_2'&L\\m_1'&m_2'&-M\end{array}\right)\left(C_{l_1'}F_{l_2'Ll_2'}+C_{l_2'}F_{l_1'Ll_2'}\right)\frac{a_{l_1'm_1'}a_{l_2'm_2'}-\langle a_{l_1'm_1'}a_{l_2'm_2'}\rangle_{\mathrm{CMB}}}{2C_{l_1'}^{tot}C_{l_2'}^{tot}},\nonumber\\
\end{eqnarray}
where
\begin{eqnarray}
\langle a_{l_1m_1}a_{l_2m_2}\rangle_{\mathrm{CMB}}&=& C^{tot}_{l_1}\delta_{l_1l_2}\delta_{m_1-m_2}(-1)^{m_2}\nonumber\\
&+&\sum_{L'M'}(-1)^{M'}\left(\begin{array}{ccc}l_1&l_2&L'\\m_1&m_2&-M'\end{array}\right)\left(C_{l_1}F_{l_2L'l_1}+C_{l_2}F_{l_1L'l_2}\right)\psi_{L'M'}.
\end{eqnarray}

\section{Estimator of the ISW-lensing bispectrum for Method II}
\label{noise_bias_method2}
Using the reconstructed lensing potential, $\hat \psi_{lm}$, and the
temperature data, one can compute the temperature-lensing cross-power
spectrum as $\hat C^{T\Psi}_l=(2l+1)^{-1}\sum^{l}\limits_{m=-l}
a_{lm}\,\hat \psi^*_{lm}$, hence the ISW-lensing bispectrum (c.f. Eq. \eqref{isw_bispectrum}):
\begin{eqnarray}
\hat b^{\mathrm{ISW-lensing}}_{l_1 l_2
 l_3}=\frac{l_2(l_2+1)+l_3(l_3+1)-l_1(l_1+1)}2 C^{TT}_{l_2}\hat
 C^{T\Psi}_{l_3}+ (5\;\mathrm{perm}), 
\end{eqnarray}
where $C^{TT}_l$ is the theoretical temperature power spectrum and $\hat
C^{T\Psi}_l=(2l+1)^{-1}\sum^{l}\limits_{m=-l} a_{lm}\,\hat \psi^*_{lm}$ 
is the temperature-lensing
cross-power spectrum estimated from the data.

However, if $\hat \psi_{lm}$ is estimated from the
temperature data themselves, $\hat \psi_{lm}$ contains a product of two
$a_{lm}$'s, and thus the ensemble 
average of $\hat C^{T\Psi}_l$ picks up the bispectrum of $a_{lm}$. In
the absence of primordial non-Gaussianity this is not an issue; however,
the presence of primordial non-Gaussianity (such as $f_{\rm NL}$)
produces a bias in $\hat C^{T\Psi}_l$. This effect needs to be taken
into account when we write down an estimator of the ISW-lensing
bispectrum. 

Using Eq. \eqref{quad_estimator_eq}, we find
\begin{eqnarray}
\langle \hat C^{T\Psi}_{l} \rangle &=& (2l+1)^{-1}\sum^{l}\limits_{m=-l} \langle a_{lm}\,\hat \psi^*_{lm} \rangle\nonumber\\
&=& \frac{N_l}{2l+1}\sum_{lm} 
\sum_{l_1'm_1'}\sum_{l_2'm_2'}\left(\begin{array}{ccc}l_1'&l_2'&l\\m_1'&m_2'&m\end{array}\right)\left(C_{l_1'}F_{l_2'll_1'}+C_{l_2'}F_{l_1'll_2'}\right)\frac{\langle
a_{l_1'm_1'}a_{l_2'm_2'}
 a_{lm}\rangle}{2C_{l_1'}^{tot}C_{l_2'}^{tot}},\label{hat_ClTpsi_mean} 
\end{eqnarray}
where we have used $\hat \psi^*_{lm}=(-1)^{m} \hat \psi_{l,-m}$.
As discussed in Sec. \ref{basics}, the expectation value of the 3-point
correlation is given by 
\begin{eqnarray}
\langle a_{l_1'm_1'}a_{l_2'm_2'} a_{lm} \rangle =\mathcal G^{m_1 m_2
 m_3}_{l_1 l_2 l_3} \left(f_{\mathrm{NL}}\,b^{\rm local}_{l'_1 l'_2 l}  + b^{\mathrm{ISW-lensing}}_{l'_1 l'_2 l} \right),\label{3alm}
\end{eqnarray}
where 
\begin{equation}
\mathcal G^{m_1 m_2 m_3}_{l_1 l_2 l_3}= \sqrt{\frac{(2l_1+1)(2l_2+1)(2l_3+1)}{4\pi}} \left(\begin{array}{ccc}l_1&l_2&l_3\\0&0&0\end{array}\right)
\left(\begin{array}{ccc}l_1&l_2&l_3\\m_1&m_2&m_3\end{array}\right).
\end{equation}
As $C^{T\Psi}_{l}$ decreases rapidly with $l$, $b^{\mathrm{ISW-lensing}}_{l'_1 l'_2 l}$ of a squeezed configuration ($l\ll l'_1\approx l'_2$) is given by
\begin{eqnarray}
b^{\mathrm{ISW-lensing}}_{l'_1 l'_2
 l}&\approx& \frac{l'_2(l'_2+1)+l(l+1)-l'_1(l'_1+1)}{2} C^{TT}_{l'_2}
 C^{T\Psi}_{l} \nonumber\\
& &+\frac{l'_1(l'_1+1)+l(l+1)-l'_1(l'_1+1)}{2} C^{TT}_{l'_2}
 C^{T\Psi}_{l}.
\label{b_isw_approx}
\end{eqnarray}
Plugging Eqs.~\eqref{3alm} and \eqref{b_isw_approx} into
Eq.~\eqref{hat_ClTpsi_mean} and then using Eqs.~\eqref{Nl_recon},
\eqref{Flll}, and \eqref{Illl}, we find
\begin{eqnarray}
\lefteqn{\langle \hat C^{T\Psi}_{l} \rangle } \label{hat_Tpsi_mean}\\
&=& \frac{N_l}{2l+1}\sum_{lm} 
\sum_{l_1'm_1'}\sum_{l_2'm_2'}\left(\begin{array}{ccc}l_1'&l_2'&l\\m_1'&m_2'&m\end{array}\right)\left(C_{l_1'}F_{l_2'll_1'}+C_{l_2'}F_{l_1'll_2'}\right)\frac{
\mathcal G^{m_1 m_2 m_3}_{l_1 l_2 l_3} \left(b^{\mathrm{ISW-lens}}_{l'_1
					l'_2 l} +
					f_{\mathrm{NL}}\,b^{\rm local}_{l'_1 l'_2 l}   \right)}{2C_{l_1'}^{tot}C_{l_2'}^{tot}}\nonumber\\
&=& \sum_l\frac{N_l}{2l+1} 
\sum_{l_1'l_2'}\sqrt{\frac{(2l'_1+1)(2l'_2+1)(2l+1)}{4\pi}}\left(\begin{array}{ccc}l_1'&l_2'&l\\0&0&0\end{array}\right)\left(C_{l_1'}F_{l_2'll_1'}+C_{l_2'}F_{l_1'll_2'}\right)\frac{
b^{\mathrm{ISW-lens}}_{l'_1 l'_2 l} + f_{\mathrm{NL}}\,b^{\rm local}_{l'_1 l'_2 l}  }{2C_{l_1'}^{tot}C_{l_2'}^{tot}}\nonumber\\
&\approx& \sum_l\frac{C^{T\Psi}_{l} N_l }{2l+1} 
\sum_{l_1'l_2'}\frac{\left(C_{l_1'}F_{l_2'll_1'}+C_{l_2'}F_{l_1'll_2'}\right)^2}{2C_{l_1'}^{tot}C_{l_2'}^{tot}}\nonumber\\
&&+\sum_l\frac{N_l}{2l+1}\sum_{l_1'l_2'}\sqrt{\frac{(2l'_1+1)(2l'_2+1)(2l+1)}{4\pi}}\left(\begin{array}{ccc}l_1'&l_2'&l\\0&0&0\end{array}\right)\left(C_{l_1'}F_{l_2'll_1'}+C_{l_2'}F_{l_1'll_2'}\right)\frac{
 f_{\mathrm{NL}}\,b^{\rm local}_{l'_1 l'_2 l}}{2C_{l_1'}^{tot}C_{l_2'}^{tot}}\nonumber\\
&=& C^{T\Psi}_{l} + \sum_l\frac{N_l}{2l+1} 
\sum_{l_1'l_2'}\sqrt{\frac{(2l'_1+1)(2l'_2+1)(2l+1)}{4\pi}}\left(\begin{array}{ccc}l_1'&l_2'&l\\0&0&0\end{array}\right)\left(C_{l_1'}F_{l_2'll_1'}+C_{l_2'}F_{l_1'll_2'}\right)\frac{
 f_{\mathrm{NL}}\,b^{\rm local}_{l'_1 l'_2 l}}{2C_{l_1'}^{tot}C_{l_2'}^{tot}},\nonumber\\
&=&C^{T\Psi}_{l} + f_{\rm NL}\,\xi_{l},
\end{eqnarray}
where 
\begin{eqnarray}
\xi_l\equiv \frac{N_l}{2l+1}\sum_{l_1'}\sum_{l_2'}\sqrt{\frac{(2l'_1+1)(2l'_2+1)(2l+1)}{4\pi}} \left(\begin{array}{ccc}l_1'&l_2'&l\\0&0&0\end{array}\right)\left(C_{l_1'}F_{l_2'll_2'}+C_{l_2'}F_{l_1'll_2'}\right)\frac{b^{\rm local}_{l'_1l'_2l}}{2C_{l_1'}^{tot}C_{l_2'}^{tot}},\nonumber
\end{eqnarray}
and, in the second line, we have used the identity of the Wigner $3j$ symbol:
\begin{equation}
\sum_{m_1 m_2}
\left(\begin{array}{ccc}l_1&l_2&l_3\\m_1&m_2&m_3\end{array}\right) \left(\begin{array}{ccc}l_1&l_2&l'_3\\m_1&m_2&m'_3\end{array}\right)=
\frac{\delta_{l_3 l'_3} \delta_{m_3 m'_3}}{2l_3+1}.
\end{equation}
Using the result above, we finally find
\begin{eqnarray}
\langle C^{TT}_{l_2} \hat C^{T\Psi}_{l_3}\rangle &=& C^{TT}_{l_2}  C^{T\Psi}_{l} + f_{\rm NL}\,C^{TT}_{l_2}\xi_{l}.
\end{eqnarray}

\section{Reconstruction noise bias in the ISW-subtracted bispectrum for
 Method III}
\label{noise_bias}
The reduced bispectrum of the ``ISW-subtracted map'' is
\begin{eqnarray}
\nonumber
 \check {b}_{l_1l_2l_3}
&=& b_{l_1l_2l_3}^{\rm local} + b_{l_1l_2l_3}^{\mathrm{ISW-lensing}}
- ({\cal G}_{l_1l_2l_3}^{m_1m_2m_3})^{-1}\left[
\langle
a_{l_1m_1}a_{l_2m_2}\frac{C_{l_3}^{T\psi}}{C_{l_3}^{\psi\psi}}\hat{\psi}_{l_3m_3}\rangle
+ (2~\mbox{perm})\right]+{\cal O}(\psi^2)\\
&=&
b_{l_1l_2l_3}^{\rm local} 
- ({\cal G}_{l_1l_2l_3}^{m_1m_2m_3})^{-1}\left[
\langle
a_{l_1m_1}a_{l_2m_2}\frac{C_{l_3}^{T\psi}}{C_{l_3}^{\psi\psi}}n_{l_3m_3}\rangle
+ (2~\mbox{perm})\right]+{\cal O}(\psi^2).\label{bispectrum_ISW_removed}
\end{eqnarray}
Here, $\hat{\psi}_{LM}$ is the reconstructed lensing potential, which is the sum of the true lensing potential $\psi_{LM}$ and reconstruction noise $n_{LM}$:
 $\hat{\psi}_{LM} = \psi_{LM} + n_{LM}$.
Using the quadratic estimator, we may reconstruct the lensing potential, $\hat{\psi}_{LM}$, by Eq. \eqref{quad_estimator_eq}, where
the reconstruction noise, $ n_{LM}$, is given by Eq. \eqref{noise_recon}.
In order to compute $\langle a_{l_1m_1}a_{l_2m_2}n_{LM}\rangle$, we need to compute $\langle a_{l_1m_1}a_{l_2m_2}a_{l_1'm_1'}a_{l_2'm_2'}\rangle$ and
$\langle a_{l_1m_1}a_{l_2m_2}\langle a_{l_1'm_1'}a_{l_2'm_2'}\rangle_{\mathrm{CMB}}\rangle$. While both of these contain the term that is linearly
proportional to the power spectrum of the lensing potential, $C_l^{\psi\psi}$, these linear terms cancel out in the difference:
\begin{eqnarray}\label{Nose_recon}
\nonumber
\lefteqn{\langle
a_{l_1m_1}a_{l_2m_2}a_{l_1'm_1'}a_{l_2'm_2'}\rangle-\langle a_{l_1m_1}a_{l_2m_2}\langle a_{l_1'm_1'}a_{l_2'm_2'}\rangle_{\mathrm{CMB}}\,\rangle
}\\
&=&  C^{tot}_{l_1}\delta_{l_1l_1'}\delta_{m_1-m_1'}(-1)^{m_1'}C^{tot}_{l_2}\delta_{l_1l_1'}\delta_{m_2-m_2'}(-1)^{m_2'}\nonumber\\
&+& C^{tot}_{l_1}\delta_{l_1l_2'}\delta_{m_1-m_2'}(-1)^{m_2'}C^{tot}_{l_2}\delta_{l_2l_1'}\delta_{m_2-m_1'}(-1)^{m_1'}\nonumber\\
&+&{\cal O}[(C^{\psi\psi})^2].
\end{eqnarray}
Therefore, we obtain
\begin{eqnarray}
\nonumber
\lefteqn{ \langle a_{l_1m_1}a_{l_2m_2}n_{LM}\rangle
}\label{aan}\\
&=&N_L\left(\begin{array}{ccc}l_1&l_2&L\\m_1&m_2&M\end{array}\right)
(C_{l_1}F_{l_2Ll_1}+C_{l_2}F_{l_1Ll_2})\nonumber
\\\nonumber
&=&
{\cal G}_{l_1l_2L}^{m_1m_2M}N_L
\left[
\frac{L(L+1)+l_1(l_1+1)-l_2(l_2+1)}{2}C_{l_1}
\right. \left.+
\frac{L(L+1)+l_2(l_2+1)-l_1(l_1+1)}{2}C_{l_2}
\right].\\
\end{eqnarray}
Here, we have used $l_1+l_2+L=\mbox{even}$ (parity invariance),
$m_1+m_2+M=0$ (triangular condition), and 
$\left(\begin{array}{ccc}l_1&l_2&L\\-m_1&-m_2&-M\end{array}\right)=\left(\begin{array}{ccc}l_1&l_2&L\\m_1&m_2&M\end{array}\right)$
(parity invariance).
Plugging Eq. \eqref{aan} into Eq. \eqref{bispectrum_ISW_removed}, we find 
\begin{equation}
  \check{b}_{l_1l_2l_3}
= b_{l_1l_2l_3}^{\rm local} + b_{l_1l_2l_3}^{noise} + {\cal O}(\psi^2),
\end{equation} 
where the reconstruction noise bias $b_{l_1l_2l_3}^{noise}$ is given by:
\begin{eqnarray}
b_{l_1l_2l_3}^{noise}
\nonumber
&=& -\frac{C_{l_3}^{T\psi}N_{l_3}}{C_{l_3}^{\psi\psi}}
\left[
\frac{l_3(l_3+1)+l_1(l_1+1)-l_2(l_2+1)}{2}C_{l_1}\right. 
\left.+\frac{l_3(l_3+1)+l_2(l_2+1)-l_1(l_1+1)}{2}C_{l_2}
\right] \\
& &+\mbox{(2 perm).}
\end{eqnarray}
The noise bias in $f_{\rm NL}$ is $\delta f_{\rm NL}= -25.3$ for the {\sl Planck} noise level, and $\delta f_{\rm NL}=-8.1$ for the cosmic-variance-limited case. 
These biases have been subtracted already in the values quoted in Table \ref{sim3}.

\bibliographystyle{JHEP}
\bibliography{/afs/mpa/home/kim/Documents/bibliography}
\end{document}